\documentclass[aps,preprint,showpacs,preprintnumbers,amsmath,amssymb,prd]{revtex4-1}
\usepackage{amssymb}
\usepackage{graphicx}

\newcommand{\be}{\begin{eqnarray}}
\newcommand{\ee}{\end{eqnarray}}

\def\ben{\begin{equation}}
\def\een{\end{equation}}
\def\bena{\begin{eqnarray}}
\def\eena{\end{eqnarray}}

\newcommand{\real}{\mbox{{\rm I\hspace{-2truemm} R}}}

\renewcommand{\v}{\rm{v}}

\def \vr {$\check{\mbox{r}}$}

\begin{document}

\title{Wormhole solutions to Ho\vr ava gravity}

\author{Marcelo \surname{Botta Cantcheff}}
\email{botta@fisica.unlp.edu.ar}

\author{Nicol\'as E. Grandi}
\email{grandi@fisica.unlp.edu.ar}

\author{Mauricio Sturla}
\email{sturla@fisica.unlp.edu.ar}

\affiliation{\it IFLP - CONICET \,\& \,Departamento de F\'\i sica - UNLP \\\it cc67, CP1900 La Plata, Argentina}

\begin{abstract}
We present wormhole solutions to Ho\vr ava non-relat\-ivistic gravity
theory in vacuum. We show that, if the parameter $\lambda$ is set to
one, transversable wormholes connecting two asymptotically de Sitter
or anti-de Sitter regions exist. In the case of arbitrary $\lambda$,
the asymptotic regions have a more complicated metric with  constant
curvature. We also show that, when the detailed balance condition is
violated softly, tranversable and asymptotically Minkowski, de Sitter
or anti-de Sitter wormholes exist.
\end{abstract}
\pacs{04.60.-m, 04.70.-s, 04.70.Bw}
\maketitle
\section{Introduction}
\subsection{Wormholes}
In general relativity, a wormhole solution is often defined as a
solution of Einstein equations with non trivial topology that
interpolates between two asymptotically flat spacetimes. A slightly
more general definition replaces the asymptotically flat regions by
other vacua as AdS or dS spacetimes, modeling gravitational
solitons. A wormhole is said to be \emph{traversable} whenever matter can
travel from one asymptotic region to the other by passing through the
throat. The classical example of a wormhole is the Einstein-Rosen
bridge \cite{einsteinrosen} which combines models of a Schwarzschild
black hole and a white hole, and it is not traversable. There are no
other traversable wormhole solutions in standard gravity in vacuum, nor
in presence of physically acceptable matter sources (see
e.g. \cite{exotic}-\cite{Morris:1988cz}).

It is a widely accepted point of view that a correct theory of quantum
gravity will modify our understanding of spacetime structure at small
scales. In particular, it is generally assumed that General Relativity
is a theory that describes large scales, and that at the UV it should
be modified by quantum effects. In this context, it seems natural to
wonder whether small wormholes in vacuum could be supported by such
quantum corrections, without requiring the addition of any exotic
matter. Indeed, in theories that correct General Relativity at short distances, like Einstein-Gauss-Bonnet, such wormholes have been shown to exist \cite{tongas}-\cite{Dotti:2007az}. They would be created at small scales, and would be invisible at the scales at which Einstein gravity is recovered.
\subsection{{Ho\vr ava-Lifshitz gravity}}

Recently, a power counting renormalizable non-relativistic theory of
gravity was proposed by Ho\vr ava \cite{horava}.  Since then, a lot of
work was made on the subject. Formal developments were presented in \cite{Calcagni:2009qw}-
\cite{gustavo}, some spherically symmetric solutions were presented in \cite{pope}-\cite{Colgain:2009fe}, toroidal solutions were found in \cite{Ghodsi:2009rv}, gravitational waves were studied in \cite{Nastase:2009nk},\cite{Takahashi:2009wc}, cosmological implications were investigated in \cite{Calcagni:2009ar}-\cite{Gao:2009bx},
and interesting features of field theory in curved space and black hole physics were presented in \cite{Myung:2009dc}-\cite{Chen:2009ka}.

~
~

 A state of the theory
is defined by a four-dimensional manifold ${\cal M}$ equip\-ped with a
three dimensional foliation ${\cal F}$, with a pseudo-Riemannian structure
defined by an Euclidean three dimensional metric in each slice of the
foliation $g_{ij}(\vec x,t)$, a shift vector $N^i(\vec x,t)$ and a
lapse function $N(\vec x,t)$. This structure can be encoded in the
ADM-decomposed metric
\be
ds^2=-N^2(\vec x,t)dt^2 + g_{ij}(\vec x,t)\left(dx^i+N^i(\vec
x,t)dt\right)\left(dx^j+N^j(\vec x,t)dt\right) \,.
\ee
The dynamics for the set $({\cal M},{\cal F},g_{ij},N_i,N)$ is defined
as being gauge invariant with respect to foliation-preserving
diffeomorfisms, and having a UV fixed point at $z=3$, where $z$ is
defined as the scaling dimension of time as compared to that of space
directions $[\vec x]=-1,[t]=-z$. This choice leads to power counting
renormalizability of the theory in the UV. To the resulting action one
may add relevant deformations by operators of lower dimensions, that
lead the theory to a IR fixed point with $z=1$, in which symmetry
between space and time is restored, and thus a general relativistic
theory may emerge \footnote{Although in principle, the model could be
  generalized to foliations where the hypersurfaces are not spacelike
  \cite{botta}}, \footnote{While this paper was in preparation,
  important objections were raised to the possibility of Ho\vr ava
  gravity would indeed flow to General Relativity, due to the presence
  of an additional degree of freedom associated to foliation violating
  diffeomorfisms, that is not gauged in Ho\vr ava theory as it is in
  General Relativity \cite{gustavo}}. This is close in spirit to the
idea of recovering four-dimensional Einstein gravity as a sort of
macroscopic limit ($z=1$) of a collection of three dimensional
theories \cite{botta} whose quantization is better understood
\cite{witten}.

In order to have a control on the number of terms arising as possible
potential terms, one may impose the so-called detailed balance
condition: the potential term in the action for 3+1 dimensional
non-relativistic gravity is built from the square of the functional
derivative of a suitable action for Euclidean three-dimensional
gravity (here three-dimensional indices are contracted with the
inverse De Witt metric). Condensed matter experience on this kind of
construction tells us that the higher dimensional theory satisfying
the detailed balance condition inherits the quantum properties of the
lower dimensional one. It has to be noted that the construction still
works even when detailed balance condition is broken softly, in the
sense of adding relevant operators of dimension lower than that of the operators appearing at the short distance fixed
point $z=3$. In the UV, the theory still satisfies detailed balance,
while in the IR, the theory still flows to a $z=1$ fixed point.

We won't go through the above described steps in more detail, but state
the resulting action together with the deformations that will be
relevant to our purposes. The interested reader can  refer to the original
paper \cite{horava}. The action for non-relativistic gravity
satisfying the detailed balance condition can be written as
\be
S=\int \sqrt{g}N\left({\cal L}_0+{\cal L}_1\right)\,,
\ee
where we have defined the Lagrangians
\bena
{\cal L}_0
&=&
 \frac{2}{\kappa^2}(K_{ij}K^{ij} - \lambda K^2)
 +
 \frac{\kappa^2\mu^2(\Lambda_W R -3\Lambda_W^2)}{8(1-3\lambda)}
\,,
\nonumber\\
{\cal L}_1
&=&
\frac{\kappa^2\mu^2 (1-4\lambda)}{32(1-3\lambda)}R^2
-
\frac{\kappa^2}{2w^4}
\left(C_{ij} -\frac{\mu w^2}{2}R_{ij}\right)
\left(C^{ij} -\frac{\mu w^2}{2}R^{ij}\right)
\,.
\eena
Here $\kappa$, $\lambda$, $\mu$ and $w$ are arbitrary couplings. The
dynamics in the infrared is controlled by ${\cal L}_0$ and then, if
$\lambda=1$, general relativity is recovered. On the other hand, in the
UV the terms in ${\cal L}_1$ become dominant, and the anisotropy
between space and time is explicit.

To define soft violations of the detailed balance condition, we may
add to the above Lagrangians a new term with the form
\bena
{\cal L}_2 &=& \frac{\kappa^2\mu^6}{8(1-3\lambda)}\,R\,,
\eena
or we can distort the relative weight of ${\cal L}_0$ and ${\cal
  L}_1$. For our purposes, it will be enough to write the deformed
action as
\be
S &=& \int dtd^3{\bf x}\,
({\cal L}_0 + (1-\epsilon_1^2){\cal L}_1)+\epsilon_2^2 {\cal L}_2\,,
\label{lagrangian-general}
\ee
the detailed balance condition is recovered for $\epsilon_1=\epsilon_2=0$. More general soft violations are also possible.

~

Since Ho\vr ava-Lifshitz theory realizes a field theory model for
quantum gravity that is power counting renormalizable in the UV, one
might expect that its classical solutions differ from those of General
Relativity at small scales, providing a mechanism to avoid
singularities. For example it may have vacuum wormholes
of the kind described above, only perceivable at microscopic scales
and hidden at large scales. The purpose of this paper is to show
that exact solutions with this behavior exist in the Ho\vr ava model.

\section{Constructing solutions with reflexion symmetry}
\label{construction}
We are going to focus our work on the simplest case of static
spherically symmetric wormholes. Under this assumption, the pseudo-Riemannian
structure of Ho\vr ava gravity can be written in terms of the metric
\be ds^2 = -\tilde N^2(\rho) dt^2 + \frac 1 {\tilde f(\rho)} d\rho^2 +
(r_o +\rho^2)^2 d \Omega_2^2\,, \ \ \ \ \ \ \ -\infty<\rho<\infty
\,.
\label{anz0}
\ee
This metric has two asymptotic regions $\rho\to\pm\infty$ connected at
$\rho=0$ by a $S_2$ throat of minimal radius $r_o$. We will also
require the additional condition of $Z_2$ reflection symmetry with
respect to the throat of the wormhole, namely
\be
\tilde
f(\rho)=\tilde f(-\rho)\ \ \ \ \mbox{and}\ \ \ \ \tilde
N^2(\rho)=\tilde N^2(-\rho)\,.
\label{z2}
\ee
As we will see, this last condition simplifies the construction of
smooth wormhole solutions in vacuum.

%

We will consider the simplest way of constructing wormhole solutions proposed by Einstein and Rosen \cite{einsteinrosen}. 
Let us first take a static spherically
symmetric solution written in the standard form
\be ds^2 = -N^2(r) dt^2 + \frac 1 {f(r)} dr^2 + r^2 d
\Omega_2^2\,,\ \ \ \ \ \ \ \ \ 0< r<\infty\,.
\label{metric}
\ee
This may or may not represent a wormhole, and it may have
singularities and/or event horizons. In the present paper we will
concentrate in the solutions of this kind presented in
\cite{pope,alex}. Such solutions are defined by the set
\bena && S\ \ \equiv \ \ ({\cal M}, f(r),
N(r))\,,\ \ \ \ \ \ \ \ \ \ \ \ \ \ \ \ \ \ 0<r<\infty\,.
\eena
Since $N_i(r)=0$ we have omitted explicit mention to the shift vector.
The foliation ${\cal F}$ is implicit in the form of the metric
(\ref{metric}) as being defined by constant $t$ surfaces. We will take
the restriction of ${\cal M}$ to ${\cal M}^+$ whose boundary is a
sphere of radius $r_o$, namely
\bena &&S^+\;\equiv\ \ ({\cal M}^+,f^+(r),
N^+(r))\,,\ \ \ \ \ \ \ \ \ \ \ \ \ r_o< r<\infty\,.  \eena
where $f^+(r)$ and $N^+(r)$ stand for $f(r)$ and $N(r)$ in the restricted domain $r>r_o$. By defining the new variable $\rho$ through  $r\equiv r_o + \rho^2 \, , \,\,\,\rho\in
\real^+_0 $ one may write
\bena &&\!\!S^+\;=\ \ ({\cal M}^+, \tilde f^+(\rho), \tilde
N^+(\rho))\,,\ \ \ \ \ \ \ \ \ \ \ \ \ 0<\rho<\infty\,,  \eena
where $\tilde N^+(\rho)=N^+(r_o+\rho^2)$ and $\tilde
f^+(\rho)=f^+(r_o+\rho^2)/4\rho^2$. Now one may define the reflected set coordenatized with $\rho\in\real_0^-$ as
\bena &&S^-\;\equiv\ \ ({\cal M}^-, \tilde f^-(\rho), \tilde
N^-(\rho))\,,\ \ \ \ \ \ \ \ \ \ \ \ \ - \infty<\rho<0\,,  \eena
where $\tilde N^-(\rho)=N^+(r_o+\rho^2)=\tilde N^+(-\rho)$ and $\tilde
f^-(\rho)=f^+(r_o+\rho^2)/4\rho^2=\tilde f^+(-\rho)$. Here ${\cal M}^-$ is a
copy of ${\cal M}^+$.

So the extended manifold may be finally defined by joining ${\cal
  M}^+$ and ${\cal M}^-$ at their boundaries at $\rho=0$, with the
foliation extended accordingly.
\bena
&&\tilde S\ \equiv\ \ (\tilde {\cal M}, \tilde f(\rho), \tilde N(\rho))\,,\ \ \ \ \ \ \ \ \ \ \ \ \ \  \ \ - \infty<\rho<\infty\,.
\eena
where $\tilde N(z)=N^+(r_o+\rho^2)$, $\tilde
f(z)=f^+(r_o+\rho^2)/4\rho^2$. Note that these functions satisfy
(\ref{z2}) and that the resulting metric has the form (\ref{anz0}).

Following Einstein and Rosen \cite{einsteinrosen}, if the equations of motion are singularity free for all values of the independent variables, we are dealing with a solution of the field equations which is regular in the region  $-\infty<\rho<\infty$. This corresponds to a spacetime composed by two identical parts $\rho>0$ and $\rho<0$ with two asymptotic regions $\rho\to\pm \infty$ joined by a hypersurface $\rho=0$ of minimal radius $r_o$ and topology $S^2\times \real$.

That the resulting set $\tilde S$ is a solution of the theory at any
point $\rho\neq0$ is made evident by changing variables back to
$r>r_o$, what results in the original solution $S^+$. On the other
hand, the change of variables is singular at $\rho=0$ ($r=r_o$), which implies that special attention should be paid to the equations of motion there. As shown in the Appendix, in order to get a smooth
solution of the vacuum equations when completing the space with the point $r=r_o$, it is sufficient to impose the condition
\be
f_{\,,r}(r_o)=0\,.
\label{stat}
\ee
In other words, the reflection point has to be chosen as a stationary point of the function $f(r)$. This is a sufficient condition for the extended set $\tilde S$ to be a solution of the source free equations of motion. Even if for the special case $\lambda=1$ more general solutions can be constructed (see Appendix), in the present paper we will stuck for simplicity to solutions satisfying the above condition.  

Moreover, we will show in the next section that condition (\ref{stat}) exactly fulfills the bound to avoid naked singularities through the formation of a transversable wormhole. In other words, there is a class of solutions (\ref{anz0}) that contain a naked singularity in $\lambda=1$ Ho\vr ava gravity \cite{pope}, but they precisely correspond to those that may extended to a transversable wormhole geometry according to the condition (\ref{stat}).

Let us finally discuss the causal behavior at the throat and the condition of transversability in the present framework. 
Since the starting solution (\ref{anz0}) may have event horizons, located at points
$r_h$ at which
\be
N(r_h)=0 \,,
\ee
The reflected solution will satisfy $\forall \rho: \tilde{N}(\rho)\neq 0$ whenever the stationary radius fulfills $r_o > r_h$. On the other hand, the vanishing of the component $g_{\rho \rho}$ at $\rho=0$ is a coordinate singularity, that does not prevent traversability. This is guaranteed by the existence of a continuous choice of the future lightcone \cite{Morris:1988tu}. In our case the time-like unit vector $u^a = \tilde{N}^{-1} \nabla^a t$ will be smoothly defined among the bridge whenever  $r_o > r_h$ \footnote{Let us observe that although the equations of motion are different, in the Einstein-Rosen bridge constructed with the black hole solution, the condition (\ref{stat}) cannot be satisfied for some $r_o>r_{h}=2m$}.
Therefore to get a traversable wormhole, a sufficient condition is:
\be r_o>r_{h+}\,,
\label{transversability}
\ee
where $r_{h+}$ is the position of the outmost horizon.

\section{Wormhole solutions to Ho$\check {\mbox{R}}$ava gravity}
\subsection{Asymptotically flat wormholes}

A spherically symmetric solution with a flat asymptotic region was
found in \cite{alex} by setting $\epsilon_1=0$, $\epsilon_2=1$ and
$\lambda=1$ in (\ref{lagrangian-general}). This theory has a Minkowski
vacuum, and it is natural to look for spherically symmetric solutions
that approaches that vacuum at infinity. As it was shown there, such solutions
exist and takes the form
\be
N^2=f=1+\omega r^2-\sqrt{r(\omega^2 r^3 + 4\omega M )}\ ,
\label{flat-BH}
\ee
here $M$ is an integration constant and $\omega=16\mu^2/\kappa^2$.
This solution has two event horizons at radius
\be
r_{h\pm}= M \left(1\pm \sqrt{1-{\frac 1 {2 \omega M^2}}}\right)\ ,
\label{flat-horizon}
\ee
which disappear leaving a naked singularity whenever
\be
2 \omega M^2<1\,.
\label{nuda}
\ee

To find a wormhole solution we will apply to (\ref{flat-BH}) the above
described reflection technique. As discussed above, this entails to
look for an stationary point of the spherically symmetric solution
\be
f_{\,,r}(r_o) = 2\omega r_o -2\frac {\omega^2 r_o^3+\omega M
}{\sqrt{r_o(\omega^2 r_o^3 + 4\omega M )}} =0 \,.
\ee
By multiplying by
$\sqrt{r_o(\omega^2 r_o^3 + 4\omega M )}$, taking the square and
rewriting in terms of the variable $x=\omega^2 r^3$, this equation can
be solved, resulting in
\be r_o=
\left(\frac{M}{2\omega}\right)^{\frac13}\,,
\label{flat-radius}
\ee
thus we can build a wormhole by taking a solution of the form
(\ref{flat-BH}) with positive $M$, and reflecting it at the value of
$r_o$ given by (\ref{flat-radius}).

In terms of this, the transversability condition
(\ref{transversability}) may be written as
\be
\left(\frac{M}{2\omega}\right)^{\frac13}>M \left(1+\sqrt{1-{\frac 1 {2
\omega M^2}}}\right)\ ,
\ee
      and can be solved by writing $x =
(2\omega M^2)^{-1/3}$ and solving for $x$, getting
\be
2\omega M^2<1\,,
\ee
which is exactly (\ref{nuda}). This implies that, in order to build a
transversable wormhole, we should start with the solution that has
naked singularity. Note that after reflection the singularity
disappears, leaving us with a completely regular wormhole.

\subsection{Asymptotically (A)dS wormholes}

If the detailed balance condition is satisfied $\epsilon_1=\epsilon_2=0$,
a spherically symmetric solution with AdS asymptotic behavior was
found in \cite{pope}, for the particular case in which $\lambda=1$. They
have the form:
\be
N^2=f=1+(\beta r)^2-\alpha \sqrt{\beta r}\,,
\label{pope1}
\ee
where $\beta = \sqrt{-\Lambda}$ and $\alpha$ is a constant of integration.

\begin{figure}[ht!]
\centering
\includegraphics[clip,width=350pt]{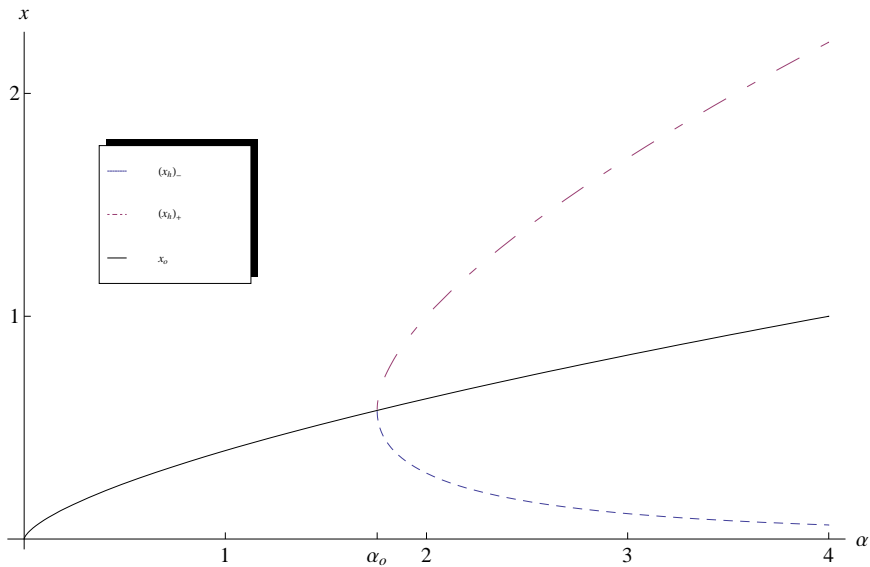}
\caption{In this plot the relation between the horizons and the
  reflection point is shown, for the case $\lambda=1$,
  $\epsilon_1=\epsilon_2=0$. As can be seen in the plot, the radius of
  the outer horizon, when it is present, is always larger than that of
  the reflection point, implying that a transversable wormhole must be
  constructed from a solution with a naked singularity.
}
\label{fig1}
\end{figure}

A  horizon will be present at the value of $r$ satisfying $N^2(r_h)=0$,
which can be rewritten as ($x=\beta r_h$)
\be
x^4 + 2 x^2 - \alpha^2 x+ 1=0 \,.
\ee
This polynomial has at most two
real roots, as can be seen from the fact that its second derivative is
always positive. Its discriminant reads
\be \Delta=\alpha^4(256 - 27
\alpha^4) \,,
\ee
and it vanishes whenever there is a double root. This
implies that the two horizons will join into a single one when $\alpha
= \pm4/3^{3/4}$. For smaller $|\alpha|$'s there is no event horizon
but a naked singularity. This can also be seen in Figure \ref{fig1}.

To find the reflection point to build our wormhole, we write the derivative as
\be
 f_{\,,r}(r_o)=2 \beta^2 r_o - \beta\frac\alpha {2\sqrt{\beta r_o}}=0\,.
\ee
This condition uniquely defines the point at which the solution must be
reflected as
\be
r_o=\frac1\beta\left(\frac{\alpha}4\right)^\frac{2}3\,.
\ee

As mentioned before, in order to ensure transversablity, the
reflection point must be at the right of the outer horizon $r_o >
r_{h+}$. This cannot be satisfied for any value of $r_o$, as can be
seen in Figure \ref{fig1}. In consequence, as it happened for the
asymptotically flat case, we should build our wormhole from the
solution without horizons $|\alpha| < 4/3^{3/4}$, which in turn
implies $r_o<1/{\sqrt{3}\beta}$.  The wormhole obtained in this way is
transversable and asymptotically (A)dS by construction.

If the detailed balance condition is relaxed to $\epsilon_1\neq0$,
$\epsilon_2=0$, a more general spherically symmetric solution with an
(A)dS asymptotic behavior has also been found with $\lambda=1$
\cite{pope}. It takes the form
\be
N^2=f= 1 +\beta^2\frac{r^2}{1-\epsilon^2_1} -
\frac{\sqrt{r(\epsilon^2_1 \beta^4 r^3+4M(1-\epsilon^2_1) \beta)}}{1 -\epsilon^2_1}
\,,
\label{AdS-BH}
\ee
where $\beta= \sqrt{-\Lambda_W}$ and $M$ a constant of integration. Despite of its similarity with (\ref{flat-BH}), this solution takes an (A)dS form asymptotically, as can be easily verified. If $\epsilon^2_1<1$ the solution has a curvature singularity at the origin. On the other hand if $\epsilon^2_1>1$, the square root becomes complex bellow a finite radius $r_c^3=4M(\epsilon_1^2-1)/\epsilon^2_1\beta^3$.

To find the event horizons, we rewrite the condition $f(r_h)=0$ as
($x=\beta r_h$)
\be
x^4+2x^2- 4 M x+(1-\epsilon^2_1)  =0\,.
\ee
Again, this polynomial in $x$ has a second derivative that is everywhere positive regardless of the values of $M$ and $\epsilon_1$, implying that it may have at most two real roots that can be identified with two event horizons. These event horizons will coincide when the roots become a single double root, namely when the discriminant vanishes
\small
\be
-256 \left(27 M^4+2 \left(9 \epsilon_1 ^2-8\right) M^2+\epsilon_1 ^4 \left(\epsilon_1 ^2-1\right)\right) =0\,,
\ee
\normalsize
or in other words when
\be
M^2_\pm=
\frac{1}{27}\left(8-9 \epsilon_1^2\pm(4-3 \epsilon_1^2)\sqrt{4-3 \epsilon_1^2}\right)\,.
\ee
Here we see that in order to have a horizon we need $\epsilon_1^2<4/3$. Under this assumption, the value of $M_-^2$ is always negative,
implying an imaginary value of $M_-$ that should be discarded. Regarding $M_+^2$,
it is always positive, determining a bound that $M^2$ should satisfy in order to have a horizon.
Such bound read $M^2>M_+^2$, or more explicitly
\be
M^2>
\frac{1}{27}\left(8-9 \epsilon_1^2\pm(4-3 \epsilon_1^2)\sqrt{4-3 \epsilon_1^2}\right)\,.
\ee
For values of $M$ satisfying this inequality, the discriminant is negative and two
horizons exists. On the other hand for $M^2<M_+^2$, there is no event horizon.

To build an asymptotically (A)dS wormhole, we reflect the above solution at the stationary point $r=r_o$ defined by
\be
f_{\,,r}(r_o) =
2\beta^2\frac{r_o}{1-\epsilon^2_1} -
2\frac{\epsilon^2_1 \beta^4 r^3_o+M(1-\epsilon^2_1) \beta}{(1 -\epsilon^2_1)\sqrt{r_o(\epsilon^2_1 \beta^4 r_o^3+4M(1-\epsilon^2_1) \beta)}}=0\,,
\ee
this  is solved by
\be
r_{o\pm} = \frac {M^{\frac13}}\beta  \left(1-\frac 2{\epsilon^2_1}
\pm
\frac1{\epsilon_1 }\sqrt{\frac 4{\epsilon^2_1 }- 3
}\right)^{\frac13}\,.
\ee
Again the condition $\epsilon_1^2<4/3$ shows up, now to ensure that a stationary point exist at which the solution can be reflected. For $1<\epsilon_1^2<4/3$, the value $r_{o+}$ is always smaller than the value $r_c$ at which the solution become complex. This implies that no wormhole can be used to cure this problem, suggesting that only small violations of the detailed balance condition $\epsilon_1^2<1$ should be accepted.

In the case $\epsilon^2<1$, it can be checked that $r_{o+}$ is never larger than $r_{h+}$. In other words, to build a transversable wormhole one should start with a solution without horizon, {\em i.e.} one satisfying $M^2<M_+^2$. Such solutions contain a naked singularity at the origin that disappears after reflection. This is analogous to what happened for cases studied in the previous sections.
It is interesting to note (see Appendix \ref{appendixA}) that a more
general class of wormhole solutions could be constructed for the
previous cases in which $\lambda=1$, without imposing the condition
$f_{,\,r}(r_o)=0$. We do not discuss it here because they do not present
a mechanism to avoid naked singularities.

\subsection{Exotic asymptotic behavior}

Spherically symmetric solutions of the action satisfying the detailed balance condition $\epsilon_1=\epsilon_2=0$ were found for arbitrary $\lambda$ in \cite{pope}.  They take the form
\bena
f&=&1+(\beta r)^2-\alpha (\beta r)^{\frac{2\lambda\pm\sqrt{6\lambda-2}}{\lambda-1}}\,,
\label{lambda-BH-f}
\\
N^2 &=& (\beta r)^{-2\frac{1+3\lambda\pm2\sqrt{6\lambda-2}}{\lambda-1}}f(r)
\label{lambda-BH-N2}\,,
\eena
where $\beta=\sqrt{-\Lambda_W}$ and $\alpha$ is a constant of integration. There are two solutions for each value of $\lambda$, defined by the choice of the sign in front of the square root.

To build a $Z_2$ symmetric solution (that we will loosely call a wormhole, even if, as we will see, its asymptotic behavior is nontrivial) we take the derivative and look for a stationary point
\bena
f_{\,,r}(r_o) &=& 2 \beta^2 r_o - \alpha \beta \frac{2\lambda\pm\sqrt{6\lambda-2}}{\lambda-1}(\beta r_o)^{\frac{2\lambda\pm\sqrt{6\lambda-2}}{\lambda-1}-1}=0\,,
\\
N^2_{\,,r}(r_o) &=&
-2\beta\frac{1+3\lambda\pm2\sqrt{6\lambda-2}}{\lambda-1}(\beta r_o)^{-2\frac{1+3\lambda\pm2\sqrt{6\lambda-2}}{\lambda-1}-1}f(r_o)=0\,.
\eena
This equations can be simplified to ($x=\beta r_o$)
\bena
2 x^2 = \alpha \frac{2\lambda\pm\sqrt{6\lambda-2}}{\lambda-1}x^{\frac{2\lambda\pm\sqrt{6\lambda-2}}{\lambda-1}}\,,
\\
\left(1+3\lambda\pm2\sqrt{6\lambda-2}\right)f(r_o)=0\,,
\eena
we see from the second equation that either the value of $\lambda$ satisfies
\be
1+3\lambda\pm2\sqrt{6\lambda-2}=0\,,
\ee
that is solved only by $\lambda=1$ when the minus sign is chosen, corresponding to the AdS case already analyzed in the previous section; or the solutions must be reflected at the horizon $f(r_o)=0$. Even if this  may result on a non-transversable wormhole, we will study the possibility of a solution of this form. In this last case equations read
\bena
\alpha \frac{2\lambda\pm\sqrt{6\lambda-2}}{\lambda-1}x^{\frac{2\lambda\pm\sqrt{6\lambda-2}}{\lambda-1}}&=&2 x^2\,,
\\
1+x^2-\alpha x^{\frac{2\lambda\pm\sqrt{6\lambda-2}}{\lambda-1}}&=&0\,,
\eena
and can be simplified to
\bena
\alpha \frac{2\lambda\pm\sqrt{6\lambda-2}}{\lambda-1}x^{\frac{2\lambda\pm\sqrt{6\lambda-2}}{\lambda-1}}&=&2 x^2\,,
\nonumber\\
\frac{2\lambda\pm\sqrt{6\lambda-2}}
{-2\mp\sqrt{6\lambda-2}}&=&x^2\,.
\label{pin}
\eena
The second equation has no solution if we choose the plus sign. On the other hand, for the minus sign, two possibilities arise from the second line, either
\bena
{2\lambda-\sqrt{6\lambda-2}}>0 \ \ \ \ \mbox{and} \ \ \ \
{-2+\sqrt{6\lambda-2}}>0\,,
\eena
or
\bena
{2\lambda-\sqrt{6\lambda-2}}<0
\ \ \ \ \mbox{and} \ \ \ \
{-2+\sqrt{6\lambda-2}}<0\,.
\eena
The first is solved by $\lambda>1$, the second instead is solved by
$1/2<\lambda <1$.

We conclude that, when $\lambda>1/2$, the second equation in (\ref{pin}) provides a value of $r_o$ at which the solution can be reflected. The first equation on the other hand, determines the value of $\alpha$ that is compatible with that $r_o$. In other words wormhole solutions can be built from the solution with the minus sign whenever $\lambda>1/2$, by reflecting  the exterior region at the horizon.

To explore the asymptotic behavior of the resulting wormhole, we take the $r\to\infty$ limit. We see that the second term in $f(r)$ in (\ref{lambda-BH-f}) will dominate whenever
\be
\frac{2\lambda-\sqrt{6\lambda-2}}{\lambda-1}<2\,,
\ee
which is always satisfied for $\lambda>1/2$. Regarding $N^2(r)$ in (\ref{lambda-BH-N2}), we see that for large $r$ it behaves as a power law with exponent
\be
n=-2\frac{1+3\lambda-2\sqrt{6\lambda-2}}{\lambda-1}+2\,,
\ee
which is positive and smaller than $4$ for $1/2<\lambda<3$ and negative for $\lambda>3$. The (A)dS case $n=2$ corresponds to $\lambda=1$ and has been studied in the previous sections.

This results in an asymptotic metric with the form
\be
ds^2 = -(\beta r)^n dt^2 + \frac{dr^2}{(\beta r)^2}+r^2d\Omega_2^2\,.
\ee
The scalar curvature is asymptotically constant and finite
\bena
R&=&\frac {\Lambda_W }2(12+4n+n^2)+\frac 2{r^2}\,,
\nonumber\\
&\to&\frac {\Lambda_W }2(12+4n+n^2)\,,
\eena
and it never vanishes as a function of $n$. As can be easily seen, the asymptotic isometry group is $SO(3)\times \real$, the first factor representing the rotations on the sphere, while the second refers to time translations.

\section{Concluding remarks}

In this work we have constructed wormhole solutions of Ho\vr ava
gravity theory in vacuum by reflection of known spherically symmetric
metrics. For both the asymptotically Minkowski or (A)dS cases, the
condition to have purely exterior solutions, namely wormholes
constructed with regions exterior to the horizon of the starting
solution, that we imposed in order to ensure transversability, forces
us to choose the parameters in the starting spherically symmetric
solutions that imply a naked singularity. After reflection the
singularity disappears, leaving us with a $Z_2$ and spherically
symmetric wormhole, where two asymptotic regions are connected through
a topologically $S^2$ neck.

Under the detailed balance condition, the $\lambda=1$ case provides a
wormhole with dS or AdS asymptotic regions. A similar solution is
obtained when the detailed balance condition is softly broken by a
deformation of the kind parameterized by $\epsilon_1$, whenever
$\epsilon_1^2<1$. In the case $\epsilon_1^2>1$ no wormhole exist and
the solution become complex at finite distance to the origin, what
suggest that only small violations of the detailed balance condition
$\epsilon_1^2<1$ should be allowed. On the other hand, if $\epsilon_2$
is turned on, the resulting asymptotic regions correspond to flat
Minkowkian spacetime. On the general $\lambda>1/2$ case, the
asymptotic regions have constant curvature, and isometry group
$SO(3)\times \real$.

The fact that a transversable wormhole can be constructed when the
starting solution has no horizon suggest the following interpretation:
if we start with value of the integration constant $\alpha$ (or $M$) for which a
horizon exists, an outside observer cannot say whether the solution represents
a microscopic wormhole hidden inside the horizon, or a black-hole with a censored singularity at the center. When varying $\alpha$ the horizon shrinks and at some point it disappears. From the macroscopic point of view, described by a relativistic theory of gravity, the singularity would become naked violating the cosmic censorship principle. But from the microscopic non-relativistic point of view, the wormhole provides a mechanism to avoid it.  Therefore, we may observe a sort of complementarity between these two mechanisms to censure the singularities.

As another related point, we may see that for instance in the solution
(\ref{pope1}) the condition to have a well defined Hawking temperature
\cite{pope} is exactly the opposite to the one that allows to form a
transversable wormhole. This is consistent with the fact that in the
last case, the degrees of freedom in both sides of the throat are
causally connected, and therefore, one should't take traces to
evaluate the asymptotic quantum amplitudes.


A remark should be made about our definition of tranversability. In this paper we have constructed  wormholes that are tranversable for particles that move according to the causal structure of the metric, namely inside the light cones. Since in the present context symmetry between space and time is broken, one may imagine particles with a more complicated causal behavior. For example a Lifshitz scalar has corrections to its dispersion relation that are quartic in momenta. Depending on the sign of the correction, the field quanta may eventually follow worldlines that travels outside the metric light cones. This may indeed enlarge the set of solutions which are microscopically transversable by such particles.

\acknowledgements{
The authors want to thank Gast\'{o}n Giribet for valuable discussions and Alex Kehagias for help and encouragement. We are also grateful to Guillermo Silva for providing a comfortable work environment during the preparation of the manuscript.}


\newpage

\section{Appendix}
\label{appendixA}
In the present Appendix we are going to show that the spherically symmetric wormhole solutions built in this paper, are in fact vacuum solutions of the equations of motion. To that end, we start from the ansatz (\ref{anz0})
\be
ds^2=-\tilde N^2(\rho)dt^2+\frac1{\tilde f(\rho)} \,d\rho^2 + (r_o+\rho^2 )^2d\Omega_2^2
\,, \ \ \ \ \ \ \ -\infty<\rho<\infty.
\label{anz}
\ee
As explained in section \ref{construction}, away from $\rho=0$ a
solution to Ho\vr ava equations with this general form can be obtained from the spherically symmetric metrics presented in \cite{pope} and \cite{alex}, by identifying $r =  r_o+\rho^2$. We will prove that, if $r_o$ is chosen to be a stationary point of $f$, the functions can be extended with continuity at the matching point $\rho=0$ (or $r=r_o$) to obtain a solution in $-\infty<\rho<\infty$, in such a way that the presence of any additional matter shell is not needed.

~

To check whether a matter shell at $\rho=0$ is required, we will first write the equations of motion in the form
\be
({\rm gravity})= ({\rm matter})\delta (\rho),
\ee
where in the left hand side we include all the contributions coming
form the Ho$\check{\rm r}$ava action, while in the right hand side we
include a $\delta$-function source describing the hypothetical matter
shell. Then we will integrate on the variable $\rho$ both sides of the above equation in the interval $(-\epsilon_a,\epsilon_b)$:
\be
\int_{-\epsilon_a}^{\epsilon_b} \,
d\rho\,({\rm gravity})=\left.({\rm matter})\right|_{\rho=0}\,.
\label{matter}
\ee
Finally we will take the $\epsilon_a,\epsilon_b \to0$ limit. Only if the resulting left hand side is non-vanishing in that limit, we will conclude that a non trivial source must contribute to the right hand side.

It is interesting to observe that for all the solutions considered in this work, the functions $N(\rho)$ and $A(\rho)\equiv 1/{f(r_o+\rho^2)}$ satisfy the following relations:
\be
N(\rho)=\frac{\beta^\omega(r_o + \rho^2)^\omega}{\sqrt{A(\rho)}},
\label{N_gen}
\ee
where $\omega=0$ and $\beta=1$ in the case $\lambda=1$, and
$\omega=-\frac{1 + 3 \lambda + 2 \sqrt{6 \lambda - 2}}{\lambda - 1}$ and $\beta$ an arbitrary constant, in the case $\lambda \neq 1$. The reader should not confuse $\omega$ and $\beta$ with the previously defined integration constants with the same name, that will not be referred in the present Appendix.

On the other hand, it is not difficult to see that, for all
the solutions we have considered, it is possible to expand $A'(\rho)$ near $\rho=0$ as follow:
\be
A'(\rho)= \textrm{c}_1 \rho +
\textrm{c}_3 \rho^3 + O[\rho^5],
\label{desA}
\ee
where $\textrm{c}_1\propto{f_{,r}(r_o)}$. This behavior at small $\rho$ is crucial in order to prove our claim.

In what follows, we compute the different contributions to
(\ref{matter}) coming from all the solutions to Horava gravity,
presented in this paper. We will evaluate the left hand side of
(\ref{matter}) using a general $A(\rho)$ with continuous first
derivative, and $N(\rho)$ with the general form expressed in (\ref{N_gen}).
\begin{itemize}
\item The contribution to the time component of (\ref{matter}) results
\be
\left.
	({\rm matter})_{N}\,
\right|_{\rho=0}
&=&
\lim_{\epsilon_a,\epsilon_b \to0 } 
\int^{\epsilon_b}_{-\epsilon_a} d\rho \, 
\left(
	\phantom{\frac12}\!\!\!\!\!
	512 \,\rho \,(r_o + \rho^2)^4 
	\left(
		\phantom{1^2}\!\!\!\!\!\!
		\epsilon_1 (2\lambda\!-\!1)
	\right.
\right. 
\nonumber \\ &&
\left.
	\left.
		+(r_o + \rho^2)^2 
		(\epsilon_0 \Lambda+\epsilon_2 \mu^4)
		\phantom{1^2}\!\!\!\!\!
	\right) 
	A(\rho)^3  
	+ 
	256 \rho (r_o + \rho^2)^4
  	\left(
  		\phantom{1^2}\!\!\!\!\!
  		\epsilon_1\!-\!2 \epsilon_1 \lambda 
	\right.
\right.
\nonumber \\ &&
\left.
	\left. 
		+(r_o+\rho^2)^2
  		(
  		\epsilon_0 \Lambda (3 (r_o + \rho^2)^2 \Lambda-2)
  		-2 \epsilon_2\mu^4
  		)
  		\phantom{1^2}\!\!\!\!\!
	\right) 
	A(\rho)^4 
\right. 
\nonumber\\ && 
\left. 
	-256 (r_o + \rho^2)^5 \epsilon_1\lambda A(\rho)\,A'(\rho) 
	-\frac{32}\rho{(r_o + \rho^2)^6 \epsilon_1
	(\lambda\!-\!1)A'(\rho)^2}
\right.
\nonumber \\ && 
\left.
	-256 (r_o + \rho^2)^4 \,A(\rho)^2 
	\left(
		\phantom{1^2}\!\!\!\!\!
  		\rho\, \epsilon_1 
  		(
  		2 \lambda\!-\!1)
	\right. 
\right. 
\nonumber\\ &&  
\left. 
	\left.
		+ (r_o + \rho^2) 
		(
		(r_o +  \rho^2)^2 (\epsilon_0 \Lambda + \epsilon_2\mu^4)
		-\epsilon_1 \lambda 
		)
		\,A'(\rho)
		\phantom{1^2}\!\!\!\!\!
	\right)
	\phantom{\frac12}\!\!\!\!\!
\right).
\label{temporal}
\ee 
Plugging the expansion (\ref{desA}) into (\ref{temporal}), we see
that the integrand is completely regular, and therefore, the temporal component of (\ref{matter}) must have a vanishing right hand side.
\item Once the integral of the regular terms is omitted since it vanishes trivially in the
limit $\epsilon_a,\epsilon_b \to0 $, the non trivial contribution to the radial ($\rho\rho$) component of (\ref{matter}) results in
\bena 
\left.(
{\rm matter})_{\rho \rho}\,
\right|_{\rho=0}
&=&
\lim_{\epsilon_a,\epsilon_b \to0 } 
\left(
	\textrm{BT}_0
  	+
  	\int^{\epsilon_b}_{-\epsilon_a} \!\!d\rho \, 
  	\left( 
  		\frac{\,32 (r_o+\rho^2)^{4+\omega}\, \beta^\omega}
  		{\rho \,\sqrt{A(\rho)}}
  		\left(
  			\left(
  				\phantom{1^1}\!\!\!\!\!
  				8 \rho (r_o + \rho^2) \epsilon_1 \,
  				( 6 \!-\! 7  \lambda) 
  				A(\rho) 
			\right. 
		\right.
	\right.
\right.
\nonumber \\ &&
\left.
	\left.
		\left.
			\left.   
				-8 \rho (r_o + \rho^2) 
				(
				-\epsilon_1\lambda 
				+(r_o + \rho^2)^2 
				(\epsilon_0 \Lambda + \epsilon_2 \mu^4)
				)
            	A(\rho)^2
            	\phantom{1^1}\!\!\!\!\!
			\right)
			\, A'(\rho)
			\phantom{\frac12}
		\right.
	\right. 
\right.
\nonumber \\ &&
\left.
	\left.
		\left.
			- {11(r_o +\rho^2)^2\,
			\epsilon_1(\lambda\!-\!1)A'(\rho)^2}
		\right)
		\phantom{\frac1{2^2}}\!\!\!\!\!\!\!
	\right)
\right)
\label{radial}
\eena
where the boundary term $\textrm{BT}_0$ reads:
\be
\textrm{BT}_0=\frac 1\rho\left(
{128 \, (r_o + \rho^2)^{6+\omega} \,\beta^\omega\,\epsilon_1 (-1+\lambda) \sqrt{A(\rho)}}
\right)A'(\rho)
|^{\epsilon_b}_{-\epsilon_a}.
\ee
Replacing the expansion (\ref{desA}) into (\ref{radial}), we see that the $\rho\rho$ component of (\ref{matter}) will have no contribution coming from the integral in (\ref{radial}). The possibility of a contribution coming from the boundary term $\textrm{BT}_0$ will be analyzed bellow.
\item The angular contribution to (\ref{matter}) results:
\begin{eqnarray}
\left.({\rm matter})_{\Omega\Omega}\right|_{\rho=0}
&=&
\lim_{\epsilon_a,\epsilon_b \to0 }
\left(
	\textrm{BT}_1 + \textrm{BT}_2 - {\textrm{BT}_3} 
	+
	\int^{\epsilon_b}_{-\epsilon_a} \!\!d \rho\,
	\frac
	{16\, (r_o+\rho^2)^{5 + \omega}\,\beta^\omega \,A'(\rho)}		
	{\rho^2 \sqrt{A(\rho)}}\,
	\times
\right.
\nonumber\\ && 
\times
\left( 
	\left.
		\frac{99}{4}\, (r_o+\rho^2)^2 \epsilon_1 \,
		(\lambda\!-\!1)\,A'(\rho)^2 
	\right. 
\right.
\nonumber\\&& 
\left.
	\left.
		\ \ -\rho\, (r_o + \rho^2)\, \epsilon_1\,( 175 -\,193 \,\lambda 
		+ 
		9 (\lambda\!-\!1)\, \omega)\, A(\rho)\,A'(\rho) 
	\right.
\right.
\nonumber \\ &&
\left. 
	\left.
	\ \	+ 2 \rho\,A(\rho)^2 \,
		\left(\phantom{1^1}\!\!\!\!
			8\rho \, \epsilon_1 \, 
			(23 \,\lambda -19  - 2 \,\lambda \,\omega)
		\right.
	\right.
\right.
\nonumber \\ && 
\left. 
	\left.
		\left.
		\ \	+9 (r_o + \rho^2) 
			(
			(r_o + \rho^2)^2 \,
			(\epsilon_0\, \Lambda + \epsilon_2 \,\mu^4)
			-\epsilon_1\, \lambda
			)
		    A'(\rho)
			\phantom{1^1}\!\!\!\!\!
		\right)	
	\phantom{\frac{2}2}\!\!\!\!\!\!\right)
\phantom{\frac{2^2}2}\!\!\!\!\!\!\!\!\right),
\label{angular}
\end{eqnarray}
where we have again omitted in the integral regular terms that do not contribute in the $\epsilon_a,\epsilon_b \to0 $ limit.  The boundary
terms that appear in (\ref{angular}) have the form
\begin{eqnarray}
\textrm{BT}_1 &=&\frac{32}{\rho^2}
\left.
{\left(r_o+\rho ^2\right){}^{7 +\omega} \,\beta^{\omega} \,\epsilon _1 (\lambda -1)\, A(\rho )^{3/2} }
A''(\rho )
\right|^{\epsilon_b}_{-\epsilon_a}
\\
~
\nonumber \\
\textrm{BT}_2 & =& -\frac{16}{\rho^3} { \,(r_o+\rho^2)^{6+\omega} \, \beta^\omega \,\sqrt{A(\rho)}\, A'(\rho)\,} 
\left(\phantom{\frac12}\!\!\!\!
	2 \epsilon_1\!
	\left(
		\phantom{1^2}\!\!\!\!\!\!\!\
		\rho^2 
		(17\, \lambda -15+ 2\, \omega 
		- 2\,\lambda \, \omega)
	\right. 
\right.
\nonumber\\&&
\left. 
	\left.
		+
        r_o (\lambda\!-\!1) 
		\phantom{1^2}\!\!\!\!\!
	\right) 
	A(\rho) 
	+  
	4 \rho^2
	\left(
		\phantom{1^2}\!\!\!\!\!
		(r_o + \rho^2)^2 \,(\epsilon_0 \,		\Lambda +
		\epsilon_2 \,\mu^4)
		-\epsilon_1 \lambda
	\right) 
	\, A(\rho)^2 
	\phantom{\frac12}\!\!\!\!
\right)
\left.
\phantom{\frac12}\!\!\!\!\!\!
\right|^{\epsilon_b}_{-\epsilon_a}
\label{bt2}
\\
~
\nonumber\\
\textrm{BT}_3 &=&\frac{152}{\rho^2 }
\left.
{ \,\left(r_o+\rho ^2\right){}^{7+\omega}  \beta^{\omega}\,\,\epsilon _1\, (\lambda -1) \,\sqrt{A(\rho )}  A'(\rho )^2}\,
\right|^{\epsilon_b}_{-\epsilon_a} .
\end{eqnarray}
Using the expansion (\ref{desA}) it is straightforward to prove that the integrand of (\ref{angular}) is regular. Therefore, if there is any nonvanishing contribution, it must come from the boundary terms.

Equation (\ref{bt2}) shows that the leading order of $\textrm{BT}_2$ near $\rho=0$ is proportional to 
\be 
\frac{A(\rho)^{3/2}\,A'(\rho)}{\rho^3}, 
\ee 
The vanishing of this term is guaranteed by
imposing the condition $\textrm{c}_1 =0$ in (\ref{desA}), which may
also be expressed as $f_{,r}(r_o)=0$.  In addition, this condition suffices to ensure the vanishing of the remaining boundary terms $\textrm{BT}_0,\textrm{BT}_1$and $\textrm{BT}_3$. In the limit
$\epsilon_a,\epsilon_b \to0$ it implies that the right hand side of eq.(\ref{angular}) is
zero, and then all possible contributions to the left hand side of the $\Omega\Omega$ component of (\ref{matter}) vanish.
\end{itemize}
The above analysis allows us to conclude that the condition
$f_{,r}(r_o)=0$ is sufficient 
to ensure that no matter shell is needed to support the wormhole solution at $\rho=0$.  Let us finally notice that, as explained in the end of Section II, for $\lambda = 1$ eq.(\ref{angular}) become trivially regular and the condition above is not necessary to obtain a regular solution.

\end{document}